\documentclass[pra]{revtex4}
\usepackage{amsmath}
\usepackage{amssymb}
\usepackage{latexsym}
\usepackage{graphicx}
\usepackage{gensymb}
\usepackage{bm}

\def\DT{D_{\rm T}}
\def\DT{D}

\def\eqnn#1{Eq.~(\ref{eq:#1})}
\def\eqnb#1{(\ref{eq:#1})}

\def\figno#1{Fig.~\ref{fig:#1}}

\def\vev#1{\left\langle #1\right\rangle}

\begin{document}
\title{Expectation values in the random walk theory  and the diffusion equation}
\author{Kenichiro   Aoki and Takahisa  Mitsui}
\affiliation{Research and Education Center for Natural Sciences and
  Dept. of Physics, Hiyoshi, Keio University, Yokohama 223--8521,
  Japan}
\begin{abstract}
  The relation between the expectation values computed in the random
  walk theory, and the heat kernel method for the diffusion equation is
  explained concretely. The random walk is also realized by
  simulations and their statistical uncertainties are analyzed.
\end{abstract}
\maketitle
\par

Random walk is a simple model with applications not only in various
areas of physics but in other disciplines, such as astronomy,
biology, chemistry, and finance\cite{rwChandra, Kampen,rwBN}. Due to its
utility and versatility, the random walk theory is an integral
component of standard statistical physics courses and also provides a model
for Brownian motion, often observed in student laboratories. While the
random walk depicts the motion of a single particle, it can serve as a
microscopic model for diffusion, which is often described by the
diffusion equation in a continuum theory. One manifest link between
these two pictures is that the Fokker-Planck equation for the
probability distribution of a random walk reduces to the diffusion
equation in the continuum limit\cite{rwBN}. To study the behavior of
physical quantities in the random walk theory, one needs to compute
their expectation values. However, how the expectation values computed
using random walks relate to the diffusion equation does not seem to
be stated or explained explicitly in the literature, to our
knowledge. In this note, we explicitly derive this relation and
illustrate this with concrete examples using simulations. Simulations
also provide us a way of realizing the physics picture intuitively, as
well as an opportunity to understand discretization and statistical
uncertainties.

In a diffusion process, the concentration $c(x,t)$ of matter under
study (such as solute in a solution) obeys the diffusion equation,
\begin{equation}
  \label{eq:diffEq}
  {\partial c(x,t)    \over \partial t} =\DT {\partial^2
      c(x,t) \over \partial x^2}
    \quad,
\end{equation}
where $\DT$ is the diffusion constant\cite{Gardiner}.  The equation is also
applicable to a wide variety of situations, including thermal
diffusion which is not diffusion of matter, per se.  Given an initial
concentration $c(x,t_0)$ at time $t_0$, the concentration, $c(x,t)$,
at time $t(>t_0)$ can be computed using the heat kernel (or Green's
function) $G(x,t; x',t')$, as
\begin{equation}
  \label{eq:exp}
  c(x,t)=\int _{-\infty}^{\infty}dx'\,G(x,t; x',t_0)c(x',t_0)\quad,\quad
  G(x,t;x',t')={  e^{-(x-x')^2/\left[4\DT (t-t')\right]}
    \over\left[4\pi \DT (t-t')\right]^{1/2}}
\quad.
\end{equation}
The expectation value of an arbitrary function, $f(x)$, of $x$ at
time $t$ can be obtained as
\begin{equation}
  \label{eq:expf}
  \vev{f(x)}=
  \int _{-\infty}^{\infty}dx\,  f(x)c(x,t)
  =\int _{-\infty}^{\infty}dx\int _{-\infty}^{\infty}dx'\,
  f(x)G(x,t; x',t_0)c(x',t_0)\quad,
\end{equation}
where $\vev{\cdots}$ denotes the expectation value.

A random walk is described by a Wiener process\cite{Kampen,rwBN,Gardiner},
\begin{equation}
    \label{eq:randomWalk}
    x(t)=x_0+\varDelta x(t),\quad 
    \varDelta x(t)\equiv \int_{t_0}^tdt'\,\xi(t')\quad,
     \quad    x_0=x(t_0)\quad,
\end{equation}
where the random variable $\xi(t)$ satisfies
\begin{equation}
    \label{eq:Wiener}
    \vev{\xi(t)\xi(t')}=2D\delta(t-t')\quad.
\end{equation}
The probability distribution $P(\varDelta x)$ for $\varDelta x$ is a
Gaussian, 
\begin{equation}
  \label{eq:ave}
  P(\varDelta x)\,dx =
  {e^{-\varDelta x^2/\left[4\DT (t-t_0)\right]}
    \over\left[4\pi \DT (t-t_0)\right]^{1/2}}\,dx
  \quad.
\end{equation}
Above, we used $\vev{\varDelta x^2}=2\DT(t-t_0)$, which follows from
\eqnn{randomWalk}, \eqnb{Wiener} and suppressed the argument of
$\varDelta x$ to avoid cumbersome notation.  The expectation value
$\vev{f(\varDelta x)}$ at $t$ is
\begin{equation}
  \label{eq:expRW}
  \vev{f(\varDelta x)} =   \int f(\varDelta x)P(\varDelta x)\,dx
\end{equation}
To derive this from the heat kernel formalism, \eqnn{expf}, we
normalize the integrated concentration to unity for averaging, and note
that $\varDelta x=0$ at $t_0$, so that $c(x,t_0)=\delta (x-x_0)$. This
reproduces the above expression.

The discussion above was performed for the one-dimensional
theory,  but the theory clearly applies to arbitrary dimensions and to any
number of particles.  We note that since random walks are independent,
a different dimension is equivalent to a different particle (in one-dimensional space).

\def\picW{0.49\textwidth}
\begin{figure}[htbp]
  \centering
  \includegraphics[width=\picW,clip=true]{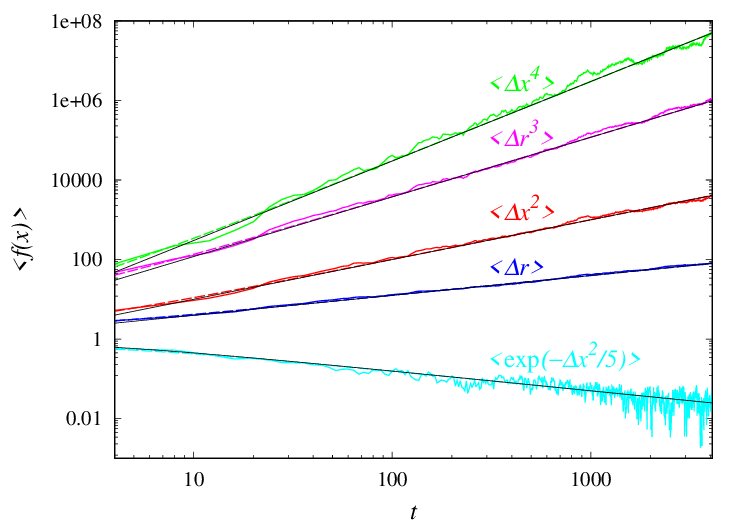}
  \caption{The expectation values of quantities (labeled in the plot)
    for $N=10^2$ (solid), $10^7$ (dashed). Their corresponding
    theoretical predictions are also shown (solid, black).}
  \label{fig:rwx}
\end{figure}
While the  stochastic method above might be advanced, depending on the
level of the students, one can also simulate random walks in
\eqnn{randomWalk} and compute the ensemble average to obtain
$\vev{f(\varDelta x)}$.  This approach is conceptually simple, and
provides an opportunity to understand the statistical errors
involved. Here, we use the simplest random walk, a symmetric walk with
unit steps, so that $2D=1$\cite{rwChandra,Kampen,rwBN,Gardiner}.  The
time and spatial steps can be rescaled to fit the situation of
interest, if desired.  One-dimensional examples for
$\vev{\varDelta x^2}, \vev{\varDelta x^4}, \vev{\exp(-\varDelta
  x^2/5)}$ and two-dimensional examples for
$\vev{\varDelta r},\vev{\varDelta r^{3}}$, where
$\varDelta r=(\varDelta x^2+\varDelta y^2)^{1/2}$, are shown for the
number of samples $N=10^2,10^7$ in \figno{rwx}. The results are
compared to the analytical results below, which follow from
\eqnn{expRW}.
\begin{equation}
  \label{eq:expAnal}
  \vev{\varDelta x^{2n+1}}=0\ ,\quad
  \vev{\varDelta x^{2n}}=(2n-1)!!(2Dt)^n\ ,\quad
  \vev{e^{-A\varDelta x^2}}=(1+4ADt)^{-1/2}\ ,\quad
  \vev {{\varDelta r^{n}}}  =(4Dt)^{n/2}\Gamma(1+n/2)
\end{equation}
Here, $\Gamma(x)$ denotes Euler's Gamma function and $n$ is a positive
integer.  The simulation results are seen to agree with \eqnn{expAnal}
well, for a large number of samples and $t\gg1$. Note that the two
dimensional examples can {\it not} be reduced to products or sums of
expectation values in one-dimensional systems, so that they are
essentially multi-dimensional in character.
The discretization uncertainties in the simulations are of the order $1/t$
relatively, which explain the slight discrepancies observed for small
$t$, and these uncertainties are consistent with the observed results.

\begin{figure}[htbp]
  \centering
  \includegraphics[width=\picW,clip=true]{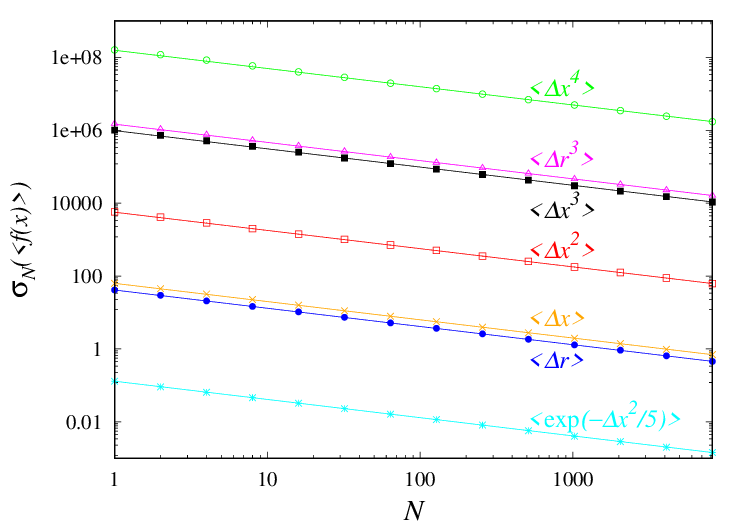}
  \caption{
    The standard deviations of the discrepancies of the
    expectation values of $\vev{\varDelta x}$ ($\times$, orange),
    $\vev{\varDelta x^2}$ ($\square$, red), 
    $\vev{\varDelta x^3}$ ($\blacksquare$, black), 
    $\vev{\varDelta x^4}$ ($\bigcirc$, green), 
    $\vev{\exp(-\varDelta x^2/5)}$ ($\ast$, cyan), 
    $\vev{\varDelta r}$ ($\bullet$, blue), 
    $\vev{\varDelta r^3}$ ($\triangle$, magenta), 
    from their theoretical predictions for $M=10^4$, which are 
    labeled by the corresponding expectation values. Their corresponding
    theoretical behaviors, \eqnn{stdExp}, are also shown and are seen to agree well. }
  \label{fig:simSTD}
\end{figure}
One can quantitatively estimate the statistical uncertainty of the
expectation value of $\vev{f(x)}$ in the simulation using its
variance, $\sigma^2_N(\vev{f(x)})=(\vev{f(x)^2}-\vev{f(x)}^2)/N$ for
$N$ samples. The variances can be computed  using \eqnn{expAnal}
to be
\begin{equation}
  \sigma^2_N\left(\vev{\varDelta
      x}\right)=2Dt/N\ ,\quad
  \sigma^2_N\left(\vev{\varDelta
      x^{2}}\right)=2(2Dt)^2/N\ ,\quad
  \sigma^2_N\left(\vev{\varDelta
      x^{3}}\right)=15(2Dt)^3/N\ ,\quad
  \sigma^2_N\left(\vev{\varDelta
      x^{4}}\right)=96(2Dt)^4/N\ ,
  \nonumber
\end{equation}
\begin{equation}
  \label{eq:stdExp}
\sigma^2_N\left(\vev{e^{-\varDelta x^2/5})}\right)=
\left[\left(1+{8\over 5}Dt\right)^{-1/2}-\left(1+{4\over 5}Dt\right)^{-1}\right]/N
\quad,
\end{equation}
\begin{equation}
  \sigma^2_N\left(\vev {\varDelta r})\right)
  =\left(4-\pi\right)Dt/N
  \quad,\quad
  \sigma^2_N\left(\vev {\varDelta r^{3}}\right)
  =(96-9\pi)(Dt)^3  /N\quad.  \nonumber
\end{equation}
In \figno{simSTD}, the standard deviations (square root of the
variances) of the differences in the simulation results from their
corresponding theoretical values at $t=4000$ are shown for
$M=10^4$ sets of $N$ samples.  They are seen to agree well with their
corresponding theoretical values for the standard deviations of the
expectation values (cf. \eqnn{stdExp}).  It should be noted that the
standard deviation of the expectation value is {\it not} the same as
that of the difference in the expectation value from its theoretical
prediction. However, the discrepancy of the
average of the expectation values from its theoretical prediction is
suppressed statistically by an additional factor of $1/\sqrt M$, which leads to the
agreement observed in \figno{simSTD}.
 \begin{acknowledgments}
 \end{acknowledgments}

\end{document}